# STUDY ON THE TRANSVERSE PAINTING DURING THE INJECTION PROCESS FOR CSNS/RCS*

M.Y. Huang[#], S. Wang, N. Huang, J. Qiu, S.Y. Xu, L.S. Huang

1：China Spallation Neutron Source, Institute of High Energy Physics,
Chinese Academy of Sciences, Dongguan, China

2：Dongguan Institute of Neutron Science, Dongguan, China

*Abstract*

For the China Spallation Neutron Source (CSNS), a combination of the $H^-$ stripping and phase space painting method is used to accumulate a high intensity beam in the Rapid Cycling Synchrotron (RCS). In this paper, firstly, the injection processes with different painting ranges and different painting methods were studied. With the codes ORBIT and MATLAB, the particle distribution and painting image were obtained. Then, the reasonable painting range which is suitable for the aperture size and magnet gap can be selected. Since the real field uniformity of BH3 and BV3 is not completely in conformity with the design requirement, the painting method and painting range also need to be selected to reduce the effects of bad field uniformity.

## INTRODUCTION

CSNS is a high power proton accelerator-based facility [1]. The accelerator consists of an 80MeV $H^-$ linac which is upgradable to 250MeV and a 1.6GeV RCS with a repetition rate of 25Hz which accumulates an 80MeV injection beam, accelerates the beam to the designed energy of 1.6GeV and extracts the high energy beam to the target. The design goal of beam power for CSNS is 100kW and capable of upgrading to 500kW [2].

For the high intensity proton accelerators, the injection with $H^-$ stripping is actually a practical method. In order to reduce the beam losses in CSNS/RCS, the phase space painting method is used for injecting the beam of small emittance from the linac into the large ring acceptance [3].

Due to the aperture size at the BH3/BV3 position becomes smaller and the real field uniformity of BH3/BV3 in some place is greater than the design requirement, the reasonable painting range and painting method for CSNS/RCS need to be studied and chosen again [4].

## PAINTING RANGE AND PARTICLE DISTRIBUTION

For CSNS/RCS, the previous aperture size at the BH3/BV3 position during the injection region was 163mm. However, due to the space at the BH3/BV3 position was so narrow, the aperture size was changed to smaller (150mm). Because of this, the transverse painting range, the painting method and the particle motion distribution during the injection process need to be studied again in detail.

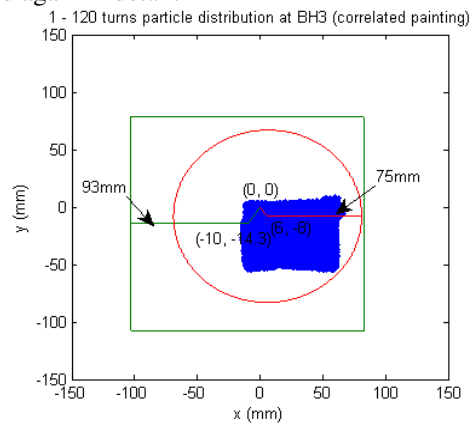

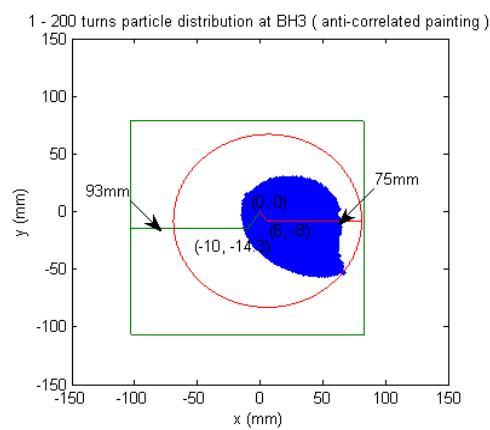

Figure 1: The particle distribution of all the 200 turns at the BH3 position for correlated painting and anti-correlated painting, respectively.

With the code ORBIT [5], the injection processes with different painting methods and painting ranges can be simulated. The painting image and particle distribution of each turn can be obtained. Then, using the code MATLAB, the particle distribution data of each turn can be analyzed and the data of all the 200 turns can be combined together. Then, it can be estimated that whether the aperture size and the magnet gap of BH3/BV3 will be suitable for the particle motion of all the 200 turns. Therefore, the reasonable painting range and painting method can be obtained.

___________________
*Work supported by National Natural Science Foundation of China
  (Project Nos. 11205185, 11175020 and 11175193)
[#]huangmy@ihep.ac.cn

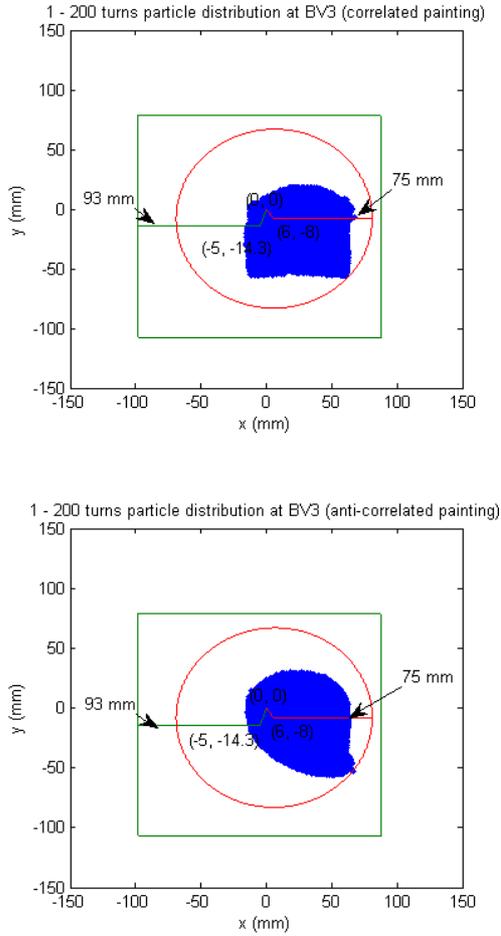

Figure 2: The particle distribution of all the 200 turns at the BV3 position for correlated painting and anti-correlated painting, respectively.

From the simulation results of the injection processes for different painting methods and painting ranges, it can be obtained that the suitable wide of the painting range were about 33mm for x-axis and 25.7mm for y-axis. Figure 1 and Figure 2 shows the particle distribution of all the 200 turns at the BH3 and BV3 position for the correlated painting and anti-correlated painting (the red circular represents the aperture size and the green square represents the magnet gap). It can be found that the aperture size and magnet gap at the BH3 and BV3 position were just suitable for the particle motion. Table 1 shows the ranges of the particle distribution at the BH3 and BV3 position for the x-axis and y-axis.

Table 1: The particle distribution ranges at the BH3/BV3 position.

|    | Painting method | x-axis (mm) | y-axis (mm) |
|----|-----------------|-------------|-------------|
| BH | Correlated      | -15<x<69    | -55<y<9     |
|    | Anti-correlated | -14<x<67    | -55<y<30    |
| BV | Correlated      | -16<x<68    | -56<y<9     |
|    | Anti-correlated | -15<x<67    | -56<y<30    |

## FIELD UNIFORMITY

The design requirement of the field uniformity for BH3 and BV3 is smaller then ±1.5%. From Figure 3 and Figure 4 which show the real field uniformity of BH3 and BV3, it can be found that the field uniformity in some places is greater than 1.5% [6]. Therefore, the effects of bad field uniformity for the real particle distribution area need to be studied in detail.

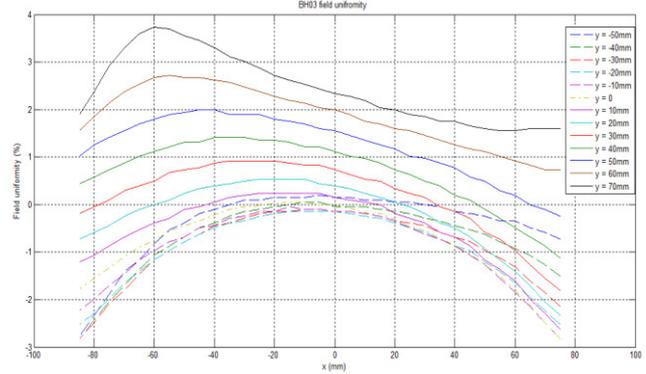

Figure 3: Field uniformity for BH3.

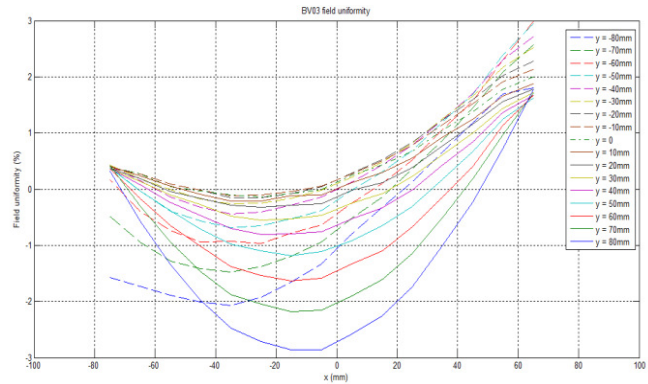

Figure 4: Field uniformity for BV3.

Compare Figure 1-2 (or Table 1) with Figure 3-4, it can be found that: for the real particle distribution area at the BH3 position, the range with bad field uniformity is about 60mm<x<70mm for x-axis whose field uniformity is between 2% and 3%; for the real particle distribution area at the BV3 position, the range with bad field uniformity is about 50mm<x<70mm for x-axis whose field uniformity is between 2% and 3%. In order to reduce the particle number in the bad field uniformity area, from Figure 1-2, compare the correlated painting with the anti-correlated painting, it can be found that the anti-correlated painting method may be more suitable for the injection process.

## CONCLUSIONS

In this paper, the injection processes for CSNS/RCS with different painting ranges and painting methods were simulated. The particle distribution and painting image of all the 200 turns during the injection process were obtained, and the reasonable painting range which is suitable for the aperture size and magnet gap were given.

The real particle distribution area at BH3/BV3 was compared with the bad field uniformity area of BH3/BV3. Then, the suitable painting method may be chosen to reduce the effects of bad field uniformity.

## ACKNOWLENDGMENTS

The authors want to thank J. Zhang, W. Kang, L. Shen, L.H. Huo, G.Z. Zhou, Y.Q. Liu and other CSNS colleagues for the discussion and consultations.

## REFERENCES

[1] S. Wang et al., Chin Phys C, 33, 1-3 (2009).
[2] J. Wei et al., Chin Phys C, 33, 1033-1042 (2009).
[3] J.Y. Tang et al., Chin Phys C, 30, 1184-1189 (2006).
[4] M.Y. Huang et al., Chin Phys C, 37, 067001 (2013).
[5] J. Gabambos et al., "ORBIT User's Manual," SNS/ORNL/AP Technical Note 011. 1999.
[6] L.H. Huo et al., "The magnetic measurement data of bump orbit injection pulse magnet for CSNS/RCS," IHEP Report (in Chinese), December 2014.